\documentclass[12pt]{iopart}
\usepackage{graphicx}
\usepackage{amssymb}
\usepackage{cite}
\begin{document}

\title{Jahn-Teller mechanism of stripe formation \\
in doped layered La$_{2-x}$Sr$_x$NiO$_4$ nickelates}

\author{      Krzysztof Ro\'sciszewski$^1$ and
              Andrzej M. Ole\'s$^{1,2}$  }

\address{$^1$ Marian Smoluchowski Institute of Physics, Jagellonian
              University, \\ Reymonta 4, PL-30059 Krak\'ow, Poland }
\address{$^2$ Max-Planck-Institut f\"ur Festk\"orperforschung, \\
              Heisenbergstrasse 1, D-70569 Stuttgart, Germany }
\ead{krzysztof.rosciszewski@uj.edu.pl; a.m.oles@fkf.mpg.de}

\date{\today}

\begin{abstract}
We introduce an effective model for $e_g$ electrons to describe
quasi-two-dimensional layered La$_{2-x}$Sr$_{x}$NiO$_4$
nickelates and study it using correlated wave functions on
$8 \times 8 $  and $6 \times 6 $ clusters. The effective Hamiltonian
includes the kinetic energy, on-site Coulomb interactions for $e_g$
electrons (intraorbital $U$ and Hund's exchange $J_H$)
and the coupling between $e_g$ electrons and Jahn-Teller distortions
(static modes). The experimental ground state phases with inhomogeneous
charge, spin and orbital order at the dopings $x=1/3$ and $x=1/2$
are reproduced very well by the model.
Although the Jahn-Teller distortions are weak, we show that they play
a crucial role and stabilize the observed cooperative charge, magnetic
and orbital order in form of a diagonal stripe phase at $x=1/3$ doping
and a checkerboard phase at $x=1/2$ doping.
\\
{\it Published in J. Phys.: Condens. Matter {\bf 23}, 265601 (2011).}

\end{abstract}



\maketitle

\section{Introduction}

The nature and origin of stripe phases in transition metal oxides
continuesly attracts considerable attention in the theory of
strongly correlated electrons. The phenomenon of stripes occurs in
doped materials and, as one of few developments in the physics of
superconducting cuprates, was discovered first in theory
\cite{Zaa89} before their existence was confirmed by experimental
observations \cite{Tra95}. In a system with dominant
electron-electron interactions novel phases with charge order in
form of a Wigner crystal, or stripe phases with nonuniform charge
distribution are expected. In the well known example of the
cuprates charge order coexists with the modulation of
antiferromagnetic (AF) order between alternating domains
\cite{strip,Ole10}. The stripe phases in the realistic models for
cuprates were obtained, {\it inter alia\/}, by calculations using
Hartree-Fock (HF) approximation \cite{Zaa89}, correlated wave
functions \cite{gora}, dynamical mean-field theory \cite{Fle00}
and slave bosons \cite{Rac06}. These studies have shown that
correlation effects play an important role in the physical
properties of stripes.

The physical origin of stripes becomes more complex in systems
with active orbital degrees of freedom, as in the doped manganites
with $e_g$ orbital degrees of freedom \cite{dag01,Feh04,Tok06},
and it was soon recognized that the mechanism of stripes has to
involve charge and magnetic order coexisting with certain type of
orbital order \cite{Hot01,theo2}. The mechanism of stripe
formation in such systems is subtle and involves directional
hopping between the orbital wave functions, in particular for
$t_{2g}$ orbital states \cite{Wro10}, that could play a role in
the properties of doped iron pnictides \cite{Jan09}.

In contrast to the stripes in cuprates \cite{strip}, the stripes in
nickelates are along diagonal direction in the square lattice
\cite{Tra94}. This by itself is puzzling and suggests that other degrees
of freedom may contribute. The properties of doped perovskite nickelates
are still not completely understood in spite of considerable effort put
forward both in theory \cite{theo2,theo0,Zaa94,theo4,theo1,theo3,theo5}
and in experiment \cite{neutrons,xrays}.
The main difficulty in the theoretical description of this class
of compounds is related to simultaneous importance of numerous
degrees of freedom. Unfortunately, all of them contribute and one
can not argue that only some types of the interactions are
essential and could be considered in a simplified approach, while
the others are of secondary importance and would be responsible
for quantitative corrections only. We argue that attempts to
consider only some interactions, for instance either including
only strong on-site Coulomb and Hund's exchange interactions and
neglecting the coupling to the lattice via the Jahn-Teller (JT)
distortions, or studying the coupling to the lattice distortions
in absence of strong local Coulomb interactions, are both not
sufficient to account for the experimental situation. The phase
situation in this class of compounds is a result of a subtle
balance between numerous factors.

On the theoretical side, the multiband models and/or approaches
based on the {\it ab initio\/} local density approximation (LDA)
computations extended by static corrections due to local Coulomb
interaction $U$ (within the LDA+$U$ scheme) \cite{theo1,theo5}, seem
to be the most realistic and complete approaches to theoretical
description of the nickel oxides with a perovskite structure.
The drawback of such approaches, however, is considerable
technical difficulty to work with doped systems especially for
arbitrary doping levels $x$ because in such a case very big unit
cells (or clusters) need to to considered.

Therefore it is quite helpful (and complementary to LDA+$U$ approaches)
to identify first all physical mechanisms which are essential in doped
nickelates and then develop a simplified but still
well motivated effective model to describe them.
In the present paper we use such an effective model (featuring
only Ni sites renormalized by the presence of surrounding oxygens)
for the description of itinerant $e_g$ electrons which includes
all essential interactions present in layered nickelates. We do
not develop new concepts here but use the model well tested before
in the manganite perovskites, including also monolayer and bilayer
systems \cite{rosc}. Further justification of this model follows
from similar microscopic approaches developed earlier to describe
doped nickelates \cite{Hot01,theo2,theo0,theo4}. With this
microscopic model we investigate the nature and type of coexisting
magnetic, orbital and charge order in the ground state when local
electron correlations and the coupling to JT distortions are both
included. In this paper we focus on quasi two-dimensional (2D)
monolayer nickelates La$_{2-x}$Sr$_{x}$NiO$_4$.

The paper is organized as follows. First, we introduce in section
2 a \emph{realistic} model for $e_g$ electrons in nickelate which
includes the electron interactions and the local potentials due to
static JT distortions. In this section we also introduce the
method to treat electron correlation effects beyond the HF
approximation, and discuss realistic values of model parameters.
The results obtained for various doping level are presented and
analysed in section 3. Next we concentrate on the role played by
the JT distortions (in section 4) and show that they are crucial
for the stability of the stripe phase at the $x=1/3$ doping.
Finally we present a short summary in section 5 as well as some
general conclusions.

\section{The microscopic model and methods}

\subsection{The effective model Hamiltonian}

We investigate strongly correlated electrons in doped monolayer
nickelates La$_{2-x}$Sr$_{1+2x}$NiO$_4$
(with doping $ 0 < x < 0.5$), using an effective model describing only
Ni sites, where the state of surrounding oxygens is included by an
effective potential at each site. (A realistic effective
Hamiltonian which acts in the subspace of low energy $e_g$ states
and the precise values of the Hamiltonian parameters
can be derived by a procedure of mapping the results of HF, or
LDA+$U$, or all-electron {\it ab initio\/} calculations obtained
within a more complete approach). A local basis at each nickel
site is given by two Wannier orbitals of $e_g$ symmetry, i.e.,
$x^2-y^2$ and $3z^2-r^2$ orbitals.

In the present study we use a Hubbard-type Hamiltonian ${\cal H}$
for two $e_g$ orbital states defined on a finite cluster:
\begin{equation}
{\cal H} = H_{\rm kin}+H_{\rm cr}+H_{\rm int}+H_{\rm spin}+H_{\rm JT} ,
\label{model}
\end{equation}
which consists of kinetic ($H_{kin}$), crystal field ($H_{\rm cr}$), on-site
Coulomb ($H_{\rm int}$), spin ($H_{\rm spin}$), and JT ($H_{\rm JT}$) terms.
The variants of this microscopic model were used to describe doped
monolayer,  bilayer and perovskite manganites \cite{rosc}, where the same
$e_g$ orbital degrees of freedom are present.

The kinetic part $H_{\rm kin}$ is expressed using
two $e_g$ orbitals:
\begin{equation}
\label{eg}
|z\rangle\equiv |(3z^2 - r^2)/\sqrt{6}\rangle, \hskip .7cm
|x\rangle\equiv |( x^2 - y^2)/\sqrt{2}\rangle.
\end{equation}
per site, with anisotropic phase dependent hopping which depends on
the orbital phases on neighbouring sites,
\begin{eqnarray}
H_{kin}&=& -\frac{1}{4} t_0 \sum_{ \{i j\} ||ab, \sigma} \Big\{
 (  3 d^{\dagger}_{i x \sigma}  d_{j x \sigma}  +
     d^{\dagger}_{i z \sigma}  d_{j z \sigma} ) \nonumber \\
& & \hskip 1.7cm \pm \sqrt{3} ( d^{\dagger}_{i x \sigma}  d_{j z \sigma}
+ d^{\dagger}_{i z\sigma}  d_{j x \sigma} ) \Big\}.
\end{eqnarray}
Here $d^{\dagger}_{i \mu \sigma}$  are creation operators for an
electron in orbital $\mu = x,z $
with spin $\sigma=\uparrow,\downarrow$ at site $i$.
The $\{ i,j\}$ runs over pairs of nearest neighbours (bonds); $\pm$
is interpreted as plus sign for $\langle ij\rangle$ being parallel
to the crystal axis $a$ and minus for $\langle ij\rangle$ being
parallel with to the axis $b$.

The kinetic energy is supplemented by the crystal field term which
describes orbital splitting
\begin{equation}
H_{\rm cr} =  \frac{1}{2} E_z \sum_{i \sigma}
( n_{iz \sigma} - n_{ix \sigma} ).
\end{equation}
For the present convention and for negative values of crystal field
parameter ($E_z < 0$) the $z$ orbital is being favored over the $x$
orbital.

The $H_{\rm int}$ and $H_{\rm spin}$ stand for an approximate form
of local Coulomb and exchange interactions for electrons within
degenerate $e_g$  orbitals used before for monolayer, bilayer and
cubic perovskite manganites \cite{rosc},
\begin{eqnarray}
H_{\rm int} &=&
  U \sum_{i \mu}  n_{i \mu \uparrow} n_{i \mu \downarrow}
 +
(U_0 - \frac{5}{2} J_H) \sum_{i}  n_{i x} n_{i z}, \\
 H_{\rm spin} &=&
 - \frac{1}{2} J_H \sum_i (n_{i x \uparrow} - n_{i x \downarrow}
)
(n_{i z \uparrow} - n_{i z \downarrow} ).
\end{eqnarray}
On-site Coulomb interaction is denoted as $U$, the Hund's
exchange interaction constant is $J_H$.
Here the spin symmetry is explicitly broken, the quantization axis is
fixed in spin space and the full Hund's exchange interactions with
SU(2) symmetry are replaced by the Ising term. However, the above form
suffices as it gives the same Hamiltonian in the HF approximation as
an exact expression for two $e_g$ orbitals \cite{Ole83}.

The simplified JT  part $H_{\rm JT}$ is:
\begin{eqnarray}
\label{HJT}
H_{\rm JT} &=& g_{JT} \sum_i \Big\{
Q_{1 i}(2 - n_{i x} - n_{i z})  + Q_{2 i} \tau_i^x
+ Q_{3 i} \tau_i^z  \Big\} \nonumber \\
&+& \frac{1}{2}\,K \sum_i \Big\{2 Q_{1 i}^2  + Q_{2 i}^2+ Q_{3 i}^2 \big)  \Big\},
\end{eqnarray}
where the pseudo-spin operators, i.e., $\{\tau_i^x,\tau_i^z\}$
operators, are defined as follows
\begin{equation}
\tau_i^x =\sum_\sigma
( d^{\dagger}_{i x \sigma}  d_{i z\sigma} +
d^{\dagger}_{i z \sigma}  d_{i x\sigma} ), \hskip .7cm
\tau_i^z =\sum_\sigma
( d^{\dagger}_{i x \sigma}  d_{i x\sigma} -
d^{\dagger}_{i z \sigma}  d_{i z\sigma} ).
\end{equation}
The Hamiltonian eq. (\ref{HJT}) includes three different static JT
modes, where $\{Q_{1 i}$,  $Q_{2 i}, Q_{3 i}\}$ denote the JT
static deformation modes of the $i-$th octahedron. (For simplicity
the harmonic constant of isotropic JT (breathing) mode $Q_1$ is
assumed to be double with respect to those corresponding to $Q_2$
and $Q_3$ unsymmetric modes as discussed in refs. \cite{dag01,
theo2}). Note that $H_{\rm JT}$ taking the above form is a
simplification: ($i$) first, some anharmonic terms were omitted
(compare for example the corresponding terms in Ref.
\cite{popovic}), and ($ii$) second, all JT distortions are treated
here as independent from each other. On the contrary, in reality
two neighbouring Ni atoms share one oxygen in between them, and
thus neighbouring JT distortions are not really independent. The
present approximation is the simplest one to describe the coupling
to the local JT modes in a model where oxygen ions are not
explicitly included (for a more detailed discussion see ref.
\cite{theo2}).

\subsection{Calculations within the Hartree-Fock approximation}

We performed extensive calculations for clusters with periodic boundary
conditions. A systematic study of increasing doping from $x=0$, through
$x=1/8$, 2/8, 3/8, up to $x=4/8$ was performed for $8\times 8$ clusters,
and was supplemented by doping $x=1/3$ for $6\times 6$ cluster.
We investigate the ground state at zero temperature ($T = 0$).

First, the calculations within the single-determinant HF
approximation were performed to determine the ground state wave
function. Also the gap between the highest occupied molecular
orbital (HOMO) and the lowest unoccupied molecular orbital (LUMO),
the HOMO-LUMO gap
\begin{equation}
\label{Delta}
\Delta \equiv E_{\rm LUMO} - E_{\rm HOMO},
\end{equation}
was extracted at this step. In the next step the HF wave function
was modified to include the electron correlations by employing
local ansatz \cite{Sto80}, see below.

Coming to details: in the first step HF computations were run starting
from one of several different initial conditions (on average a few
thousands for each set of Hamiltonian parameters are necessary as the
identification of a true energy minimum is difficult), i.e., from
predefined (some symmetry fixed, but mostly random) charge, spin and
orbital configurations and several predefined sets of classical
variables $\{ Q_{1 i},Q_{2 i},Q_{3 i} \}$. For each fixed set of
starting parameters and starting initial conditions we obtain on
convergence a new HF wave function $\vert\Psi_{\rm HF}\rangle$ which is
a candidate for a ground state wave function.
This self-consistent procedure is assumed to provide energy minimum
also with respect to the classical $\{ Q_{1 i},Q_{2 i},Q_{3 i} \}$
variables \cite{rosc}.

\subsection{Electron correlations}

After completing the HF computations for a given wave function
$|\Psi\rangle$, we performed correlation
computations (second step) which provide the total energy.
(For details see refs. \cite{gora,rosc}).
Namely the HF wave function $|\Phi_0\rangle$ was
modified to include the electron correlation effects. We used
exponential local ansatz for the correlated ground state \cite{Sto80},
\begin{equation}
|\Psi\rangle = \exp\Big(-\sum_m \eta_mO_m\Big)|\Phi_0\rangle,
\label{la}
\end{equation}
where  $\{O_m\}$
are local correlation operators. The variational parameters $\eta_m$
(four singlet and two triplet ones) 
are found by minimizing
the total energy,
\begin{equation}
E_{\rm tot}=\frac{\langle\Psi|H|\Psi\rangle}{\langle\Psi|\Psi\rangle}.
\label{etot}
\end{equation}
Here for the correlation operators we use
\begin{equation}
O_m=\sum_i \delta n_{i\mu \sigma}\delta n_{i\nu {\sigma'}}, \label{om}
\end{equation}
The sum in the above equation ensures smaller number of the variational
parameters; on the other hand the nonhomogeneous correlations are
treated in an averaged way (this procedure is somewhat similar to
calculations performed within mean-field approaches).
The symbol $\delta$ in $\delta n_{i \mu \sigma}$ indicates that
{\it only that part} of $n_{i\mu\sigma}$ operator is included which
annihilates one electron in an occupied single particle state which
belongs to the HF ground state $|\Phi_0\rangle$, and creates an electron
in one of the virtual states. The above local operators $O_m$ correspond
to the subselection of most important two electron excitations within
the {\it ab initio\/} configuration-interaction method. We note that
three, four, {\it etcetera}, electron excitations (which are omitted
here) are also important in the strongly correlation regime. However,
as yet there is no an easy way to implement them for a larger systems,
thus one is able to implement within theory only the leading part of the
correlation energy (which follows from two-particle excitations).

After obtaining the total energy for a given configuration, we
repeat all the procedure from the beginning, i.e., we take the
second set of HF initial conditions and repeat all computations to
obtain the second candidate for a ground state wave function.
Other configurations for the third, fourth, {\it etcetera}, set of
initial conditions are investigated in a similar way. Finally, the
resulting set of total energies was inspected and the lowest one
was identified as the best candidate for the true ground state. In
general, the correlations are found to be strong, in fact much
stronger than those found in the perovskite manganites
\cite{rosc}.

\subsection{Parameters of the model}

The values of the Hamiltonian parameters used below follow closely the
sets commonly employed in the literature to study doped nickelates.
For the effective $d-d$ hopping $(dd\sigma)$ we assume $t_0=0.6$ eV,
following ref. \cite{theo4}.
Local Coulomb interactions are strong, and we assume
$ 8 < U/t_0 < 12$
(the lowering of the very large atomic value on nickel is due to
an appropriate screening of the $e_g$ electrons).
Hund's exchange coupling was  $J_H =0.9 $ eV \cite{theo1,theo4}.
The crystal field values we considered were either $E_z=0$ or $E_z=-0.3$
eV, with the latter value inducing a higher electron density in $z$
orbitals, as expected for a single NiO$_2$ plane
(compare with refs. \cite{theo2,theo4}).

Very little is known about JT interactions in nickelates, therefore
the JT constant $K$ was fixed as $K = 13$ eV \AA$^{-2}$ (like in
manganite perovskites \cite{rosc}), and the coupling constant with
the lattice distortions $g_{\rm JT}$ was assumed to be in the
range 2 eV\AA${}^{-1}$ $<g_{\rm JT}<$ 3.8 eV\AA${}^{-1}$,
following ref. \cite{dag01}.

The best set of the Hamiltonian parameters which allows one to reproduce
the experimental phase situation is given in table 1. In some cases,
other parameter values were used as discussed below. However, the
parameters one should use are
constrained by the experimental observations. It is
fortunate that the obtained types of spin, orbital and charge order turn
out to be rather sensitive on the parameter values. In fact, many sets
of the Hamiltonian parameters (different than those in table 1) were also
considered in test computations but they do not generate
reasonable results (i.e., the ground states with the ordering
close to that reported in experiments).
Although full phase diagram would be very appropriate here, it is
unfortunately too expensive to compute by the present method.

\begin{table}[t!]
\caption{\label{tab:ene} Parameters of the microscopic model eq.
(\ref{model}) used in most of the performed calculations
(all electronic parameters $\{t_0,U,J_H,E_z\}$ in eV).}
\begin{center}
\begin{tabular}{cccccc}
\hline
 $t_0$ & $U$ & $J_H$ & $E_z$ & $g_{\rm JT}$ (eV \AA$^{-1}$) & $K$ (eV \AA$^{-2}$)\\
\hline
  0.6  & 7.0 &  0.9  & $-0.3$ &  3.0  &  13.0  \\ \hline
\end{tabular}
\end{center}
\end{table}

\section{Results}

The computations were repeated for many sets of the Hamiltonian parameters
and for many (up to several thousands)
charge, spin, orbital and JT distortions
(i.e., for local configurations required to start HF iterations).
Extensive calculations were required to establish the parameter values
which are realistic for doped nickelates and allow one for reproduction
of basic experimental observations. Below we present the results
obtained for the realistic parameter set as given in table 1.

\subsection{Low doping regime}

The results obtained in low doping regime are presented in figures
1-2. First we considered $x=1/8$ doping. In
La$_{2-x}$Sr$_{1+2x}$CuO$_4$ cuprates one finds at this doping
level stable stripe phases \cite{strip,gora,Fle00,Rac06}. In the
present case we observed isolated Ni ions with holes doped in $x$
orbitals, as shown in figure 1. At these ions the magnetic moments
are due to $S=1/2$ spins and are therefore lower than magnetic moments at $S=1$ sites,   
thus appearing as spin defects in                                                        
the $G$-AF phase. While both $x$ and $z$ orbitals are occupied at
undoped sites, one finds that only $z$ orbitals are occupied at
doped sites. Of course, certain delocalization of electrons
between the two sublattices takes place as well in a doped
antiferromagnet investigated here, but the above ionic picture
applies to a good approximation. In fact, the obtained phase is
insulating with a HOMO-LUMO gap of 1.86 eV.

\begin{figure}
\centerline{
\includegraphics[width=7.5cm]{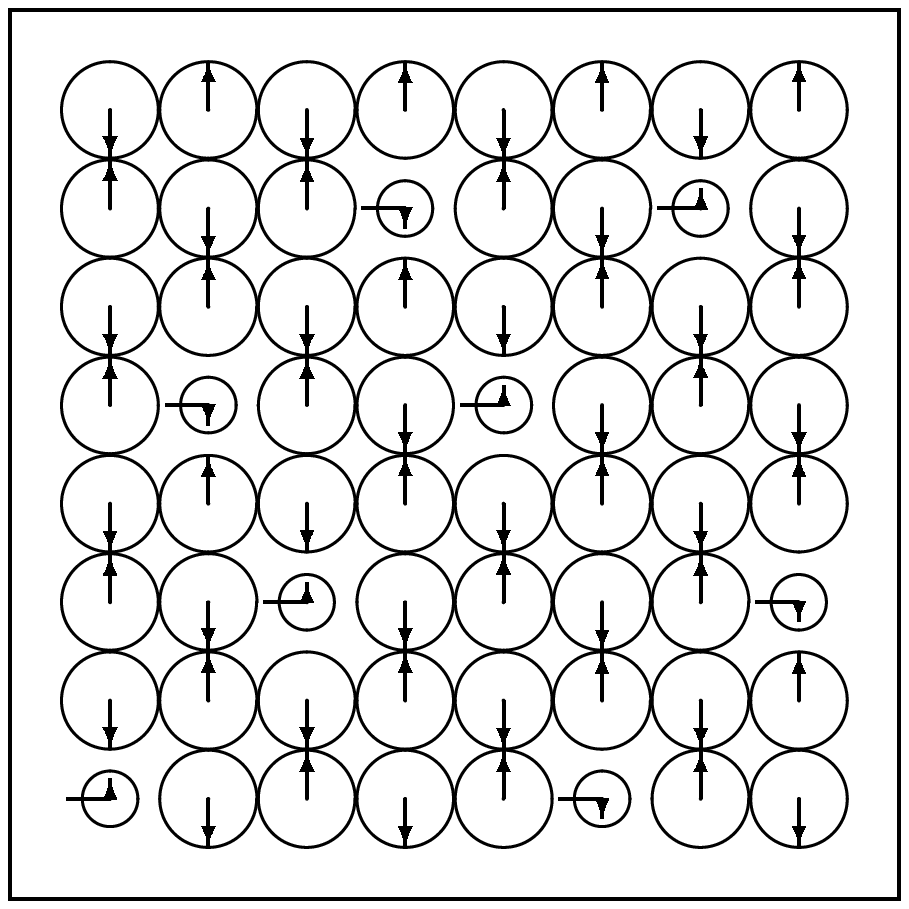}
}
\centerline{
\includegraphics[width=7.5cm]{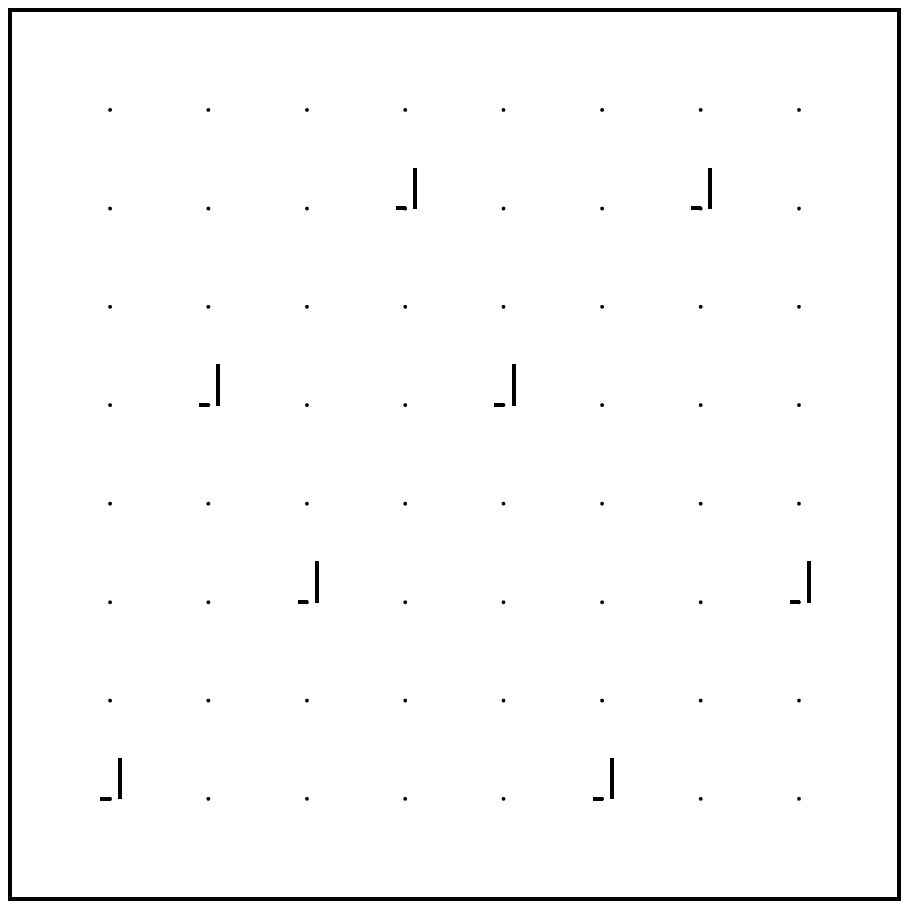}
}
\caption{%
Insulating $G$-AF (S=1) spin arrangement, with mostly uniform
charge distribution and equal electron densities within $x$ and
$z$ $e_g$ orbitals (no orbital order) at undoped sites as obtained
for $x=1/8$ doping in $8 \times 8 $ cluster. Note 8 doped holes
(charge minority sites with smaller spins) placed randomly and
with $z$ orbitals occupied by electrons. {\bf Upper panel}: At
each site the circle diameter corresponds to one-half of $e_g$
on-site total charge; the arrow length to one-half of the $e_g$
spin; the horizontal bar length to charge density difference
between $x$ and $z$ orbitals (longest bar to the right corresponds
to pure $x$, to the left to pure $z$, zero length to half by half
combination). All these values are expressed in approximate
proportionality (as generated by latex graphic package) to nearest
neighbour site-site distance which is assumed to be unity. {\bf
Lower panel}: JT distortions $Q_{2i}$ (in \AA) for the same plane
shown as bars drown slightly to the left of each site and
$Q_{3i}$, shown by bars to the right of each site. Isotropic
(breathing) mode $Q_{1i}$ in between. $Q_{2i}$ and $Q_{3i}$ bars
are artificially enlarged (by a factor of 2) for a better
visibility. The static JT distortions (finite $Q_3$ mode, the
other modes vanish) are only present at charge minority sites. }
\end{figure}

\begin{figure}[t!]
\centerline{
\includegraphics[width=7.5cm]{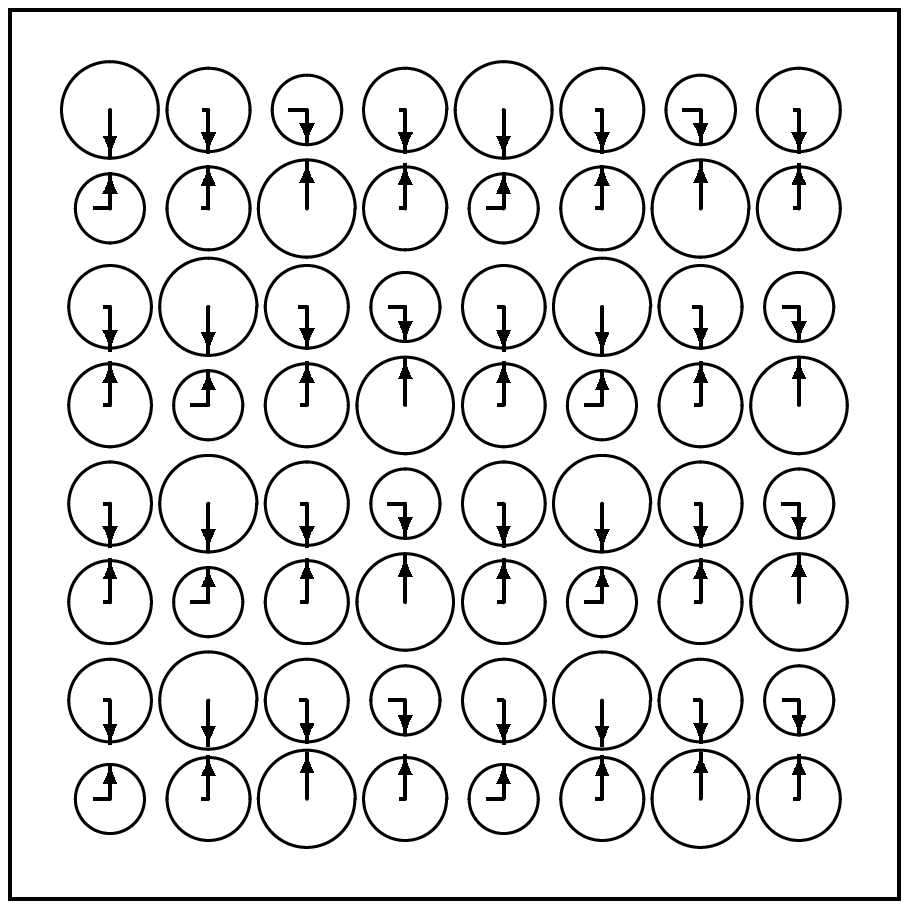}
}
\centerline{
\includegraphics[width=7.5cm]{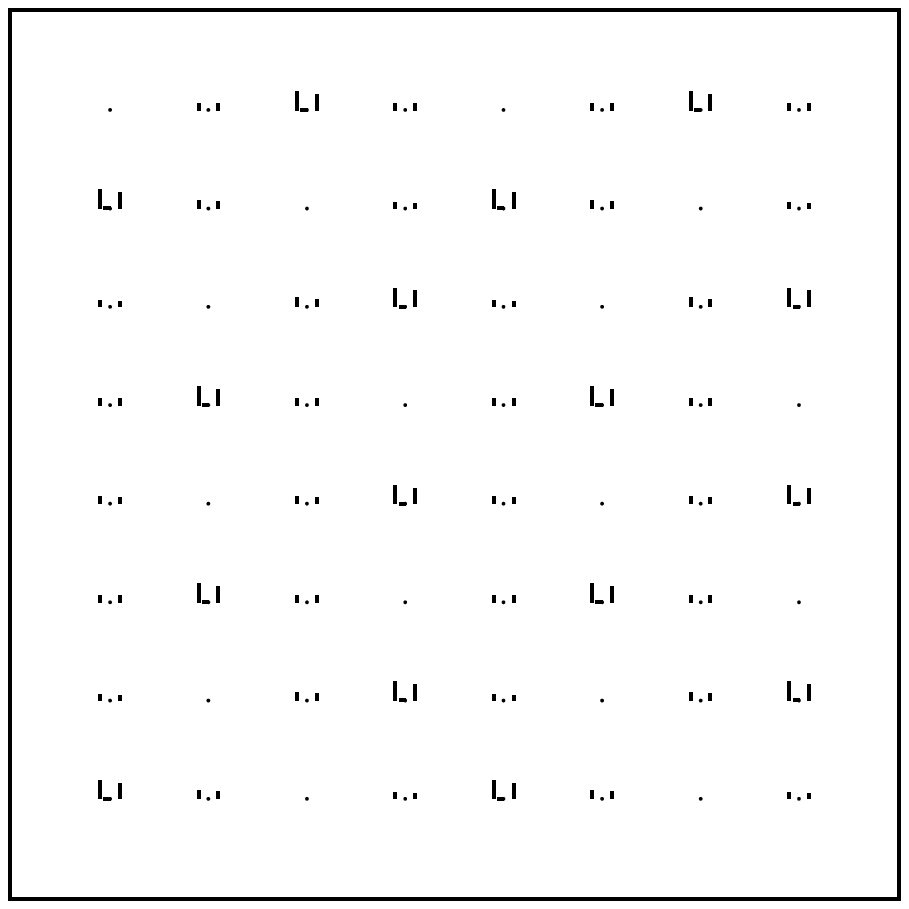}
}
\caption{%
Weakly insulating $C$-AF phase (ferromagnetic horizontal lines
coupled antiferomagnetically in vertical direction) as obtained
for the doing $x=1/4$ within the $8 \times 8 $ cluster.
Charge majority (undoped) sites have uniform charge distribution,
with both $x$ and $z$ orbitals occupied and $S=1$ spins; the
charge minority sites are occupied by $z$ electrons with spins
$S=1/2$ and show small JT distortions, $Q_2$ and $Q_3$.
Meaning of symbols and the values of parameters as in figure 1.
}
\end{figure}

Local charge defects consist of $z$ electrons occupying doped
Ni$^{3+}$ sites which couple to lattice distortions $\{Q_{i3}\}$,
as shown in the lower panel of figure 1. These distortions act as
self-trapping on the doped charges and suppress the kinetic energy
of doped holes. Therefore, this regime of doping is manifestly
different from the observations in cuprates, where JT distortions
were inactive and self-organization of doped holes in form of
horizontal stripes takes place \cite{Zaa89,Tra95,strip}.

At a higher doping $x=1/4$ the magnetic order changes to the
$C$-AF phase shown in figure 2, with relatively large magnetic
cells. This phase is weakly insulating, with the HOMO-LUMO gap
$\Delta=0.28$ eV. Horizontal ferromagnetic (FM) lines are
characterized by charge modulations between almost undoped
Ni$^{2+}$ ions and doped Ni$^{2+}$ ions. The latter doped sites
are distributed far from one another both within FM chains and
between consecutive AF horizontal lines. This may be seen as a
physical reason which prevents the occurrence of stripes at this
doping level. Furthermore, the doped sites couple actively to
local distortions, and can be identified by looking at the pattern
of JT distortions, see the lower panel in figure 2. It helps to
identify imperfect long-range order of doped holes which avoid
each other, similar to the $x=1/8$ doping considered above.

\subsection{Diagonal stripes at doping $x=1/3$}

The case of $x=1/3$ doping was investigated with $6\times 6$ cluster,
see figure 3. Here the number of doped sites is the same as the
doping, i.e., $x=1/3$, meaning that some of doped sites have to
be located at next-nearest neighbour positions. The resulting
stripe phase with diagonal lines of reduced charges (doped sites)
may be therefore seen as a modification of the $C$-AF phase
obtained at the $x=1/4$ doping by adding more doped holes
(Ni$^{3+}$ ions). Indeed, one finds that every third site in an AF
row (FM column) contains the reduced charge close to one added
hole, and these sites form a (11) stripe phase. Note that the
horizontal and vertical lines are here interchanged as compared
with figure 2, but this configuration is of course equivalent to
the one with FM horizontal lines, and AF order between them. The
large JT distortions accompany the sites with minority charges, so
the stripe pattern can be also recognized by analyzing them. It
reproduces the experimental results in refs.
\cite{neutrons,xrays}, with the long-range order coexisting
charge, spin and orbital order in form of diagonal stripes,
present at the doping $x=1/3$. Also this phase is insulating, with
a HOMO-LUMO gap $\Delta=0.57$ eV.

\begin{figure}[t!]
\centerline{
\includegraphics[width=7.5cm]{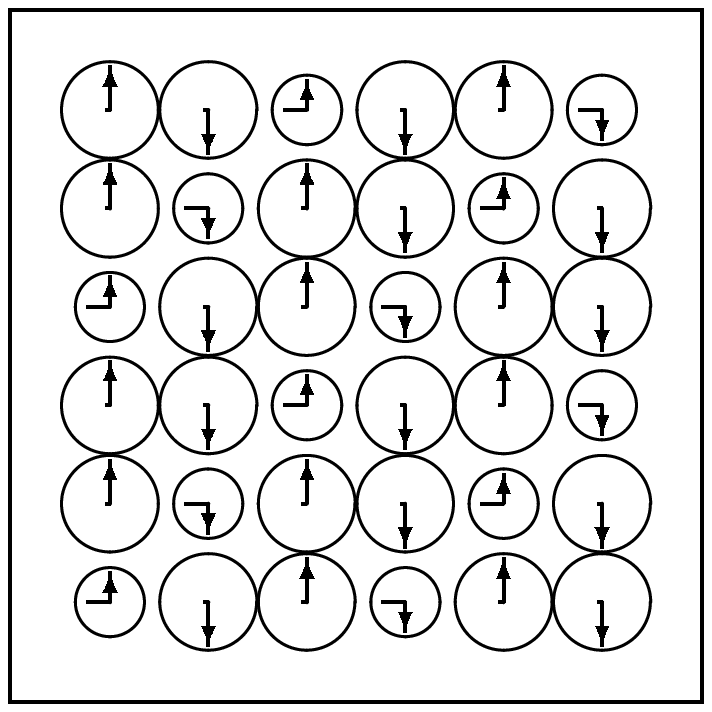}
}
\centerline{
\includegraphics[width=7.5cm]{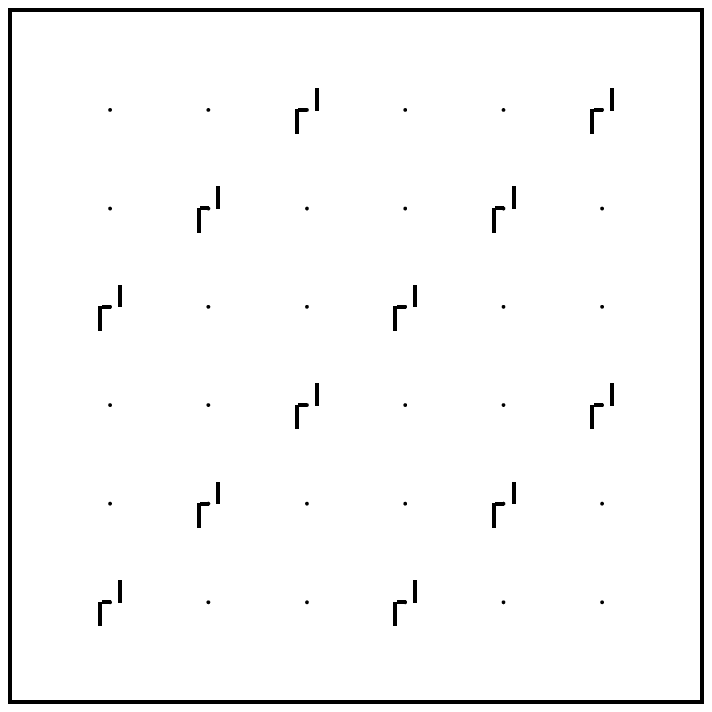}
}
\caption{%
Insulating $C$-AF phase spin arrangement (ferromagnetic vertical
lines coupled antiferomagnetically), with uniform charges on charge
majority sites (spins $S=1$ snd both $x$ and $z$ orbitals
occupied), as obtained for the $x=1/3$ doping in $6 \times 6$ cluster.
The charge minority sites form diagonal stripe boundaries. They are
occupied by $z$ electrons with spins $S=1/2$ and they show small
uniform $Q_2$ and $Q_3$ JT distortions.
Meaning of symbols and the values of parameters as in figure 1.
}
\end{figure}

We emphasize that the JT distortions play a crucial role in the
observed stripe phase at $x=1/3$ doping. We have verified that
when one switches off the JT coupling, i.e., when one puts $g_{\rm
JT}=0$ with other parameters unchanged, the stripes vanish and the
ground state becomes conducting (HOMO-LUMO gap $\Delta=0.03$ eV), see
below. In this case one finds the same magnetic phase with
horizontally arranged FM lines coupled in the vertical direction
in the $C$-AF order, with slightly nonuniform charge distribution
and slightly nonuniform orbital order ($z$ orbital occupancy is
then larger than $x$ one on most sites in the cluster).

\subsection{Large doping regime $1/3 < x \leq 1/2$}

Coming to large doping regime $x>1/3$, we have found further modifications
of the $C$-AF spin order. First of all, the next doping level considered
by us $x=3/8$ is incompatible with the stripes shown in figure 3. As
shown in figure 4, in this case the ionic picture does not apply anymore
and the electron charge is distributed uniformly. Therefore the values
of the magnetic moments are close to 1.6 $\mu_B$ per site, and the $e_g$
orbitals are occupied in a similar way at all cluster sites. Hole doping
reduces the electron density in $x$ orbitals, therefore at the present
doping level of $x=3/8$ the ratio of electron density in $x$ and $z$
orbitals is close to 1:2. Unlike for lower doping level considered
above, this electron distribution does not favor JT distortions at
particular cluster sites with minority charge, and one finds uniform
and rather small JT distortions at all the sites. Nevertheless, the
electronic structure predicts that this ground state is insulating,
having a HOMO-LUMO gap $\Delta=0.72$ eV.

\begin{figure}[t!]
\centerline{
\includegraphics[width=7.7cm]{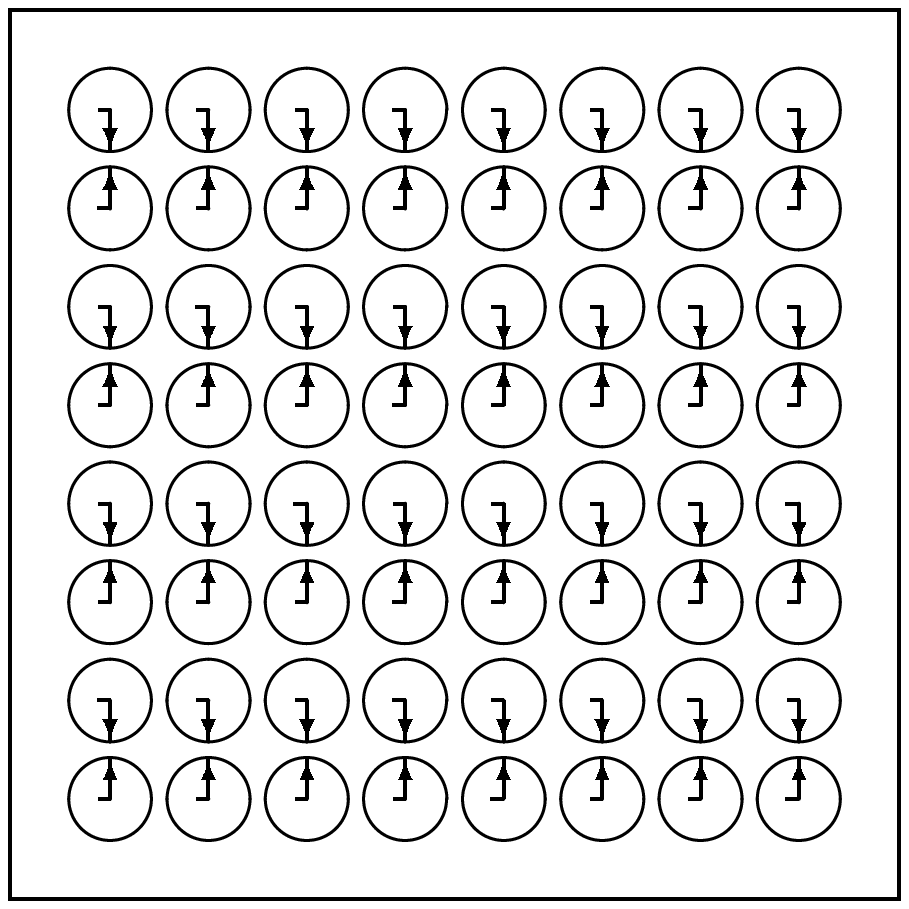}
}
\centerline{
\includegraphics[width=7.7cm]{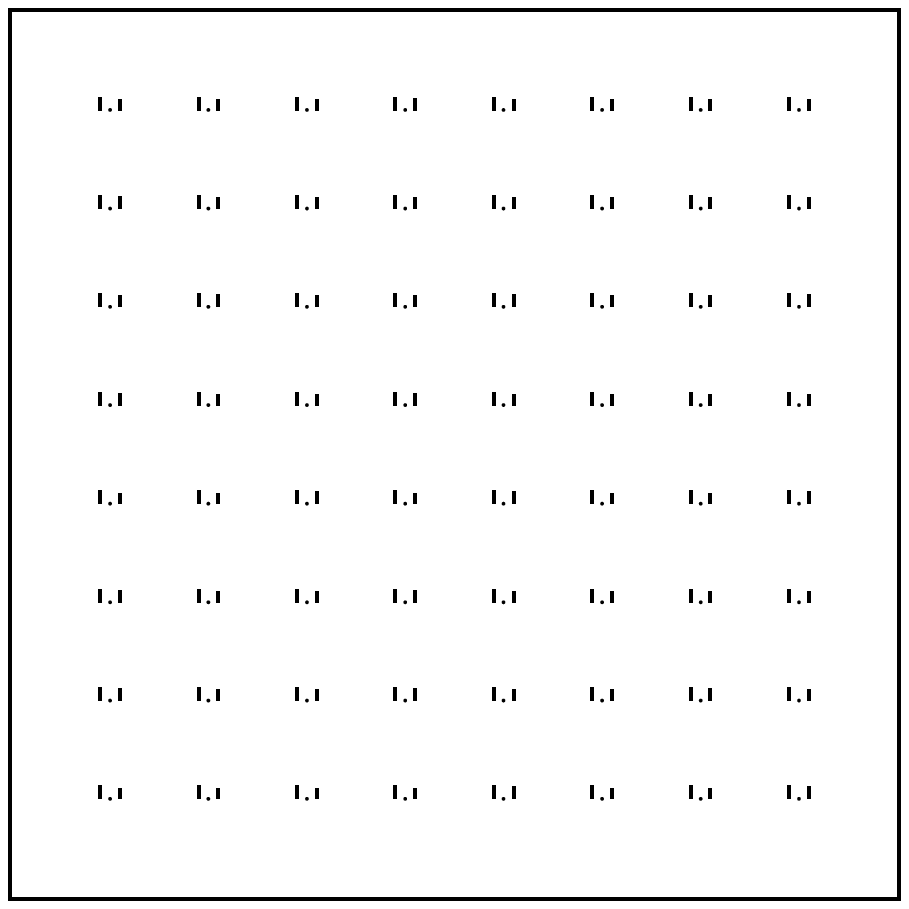}
}
\caption{%
Almost perfect $C$-AF spin arrangement (like in figures 2-3) but
with uniform charges, with about 1:2 $x$ to $z$ orbital order for
$e_g$ electrons and very small JT distortions, as obtained for
$x=3/8$ doping in $8\times 8 $ cluster. The pattern of JT
distortions is almost uniform. Meaning of symbols and the values
of parameters as in figure 1.
}
\end{figure}

The next physically interesting and experimentally much studied
doping is half filling, i.e., $x=1/2$. One finds that the $C$-AF
spin order remains unchanged, but charge modulation develops. This
doping level is compatible with a two sublattice structure in the
charge ordered state shown in figure 5. The alternating electron
charges are close to $n=2$ and $n=1$ on the majority and minority
charge sites. Therefore, the ionic picture applies again to this
doping level, and the system may be viewed as alternating charge
order, with Ni$^{2+}$ and Ni$^{3+}$ ions on the two sublattices.
This result reproduces the experimental results
\cite{neutrons,xrays}, with long-range checkerboard-like charge
order found at half doping. Similar to all other cases, also this
configuration is insulating, with a HOMO-LUMO gap $\Delta=0.91$
eV.

The charge order found for $x=1/2$ doping is closely followed by
JT distortions, shown in figure 5. Large JT distortions at charge
minority sites stabilize the checkerboard charge order in this
case. Large $\{Q_{i2}\}$ distortions stabilize the charge order,
and nonuniform electron distribution over $e_g$ orbitals is
stabilized by large $\{Q_{i3}\}$ distortions. As a result, $x$
orbitals are almost empty at charge minority sites, and the
occupancy of $z$ orbitals is close to one electron at each site.

\begin{figure}[t!]
\centerline{
\includegraphics[width=7.7cm]{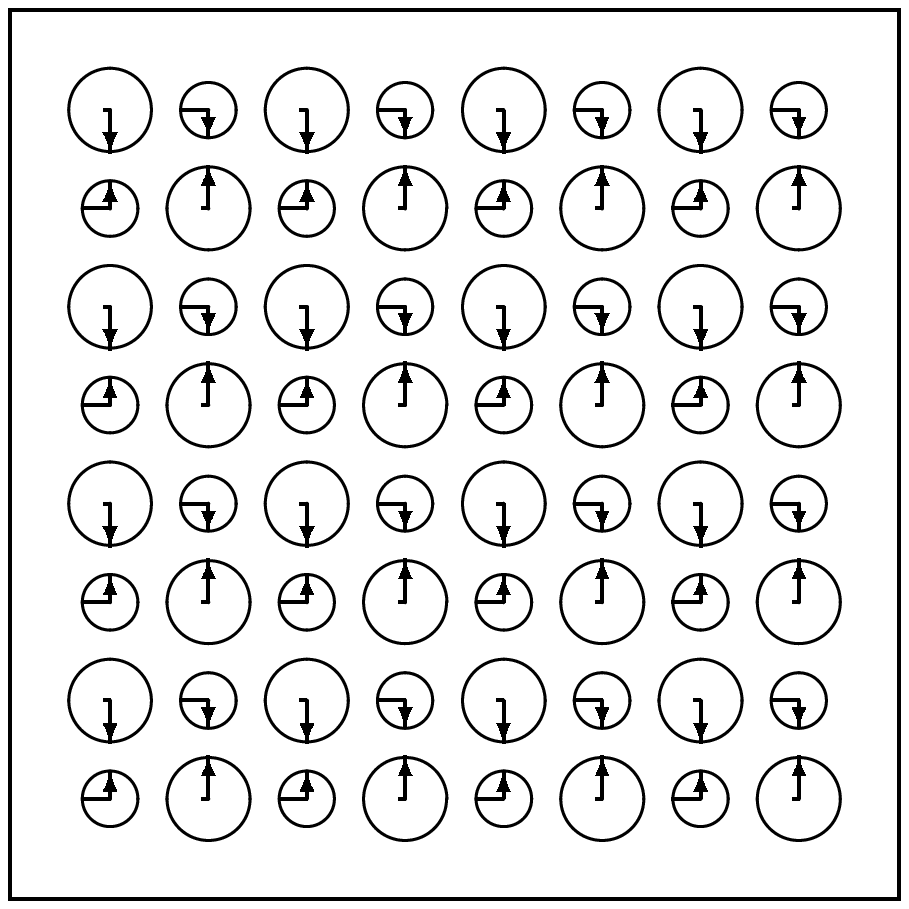}
}
\centerline{
\includegraphics[width=7.7cm]{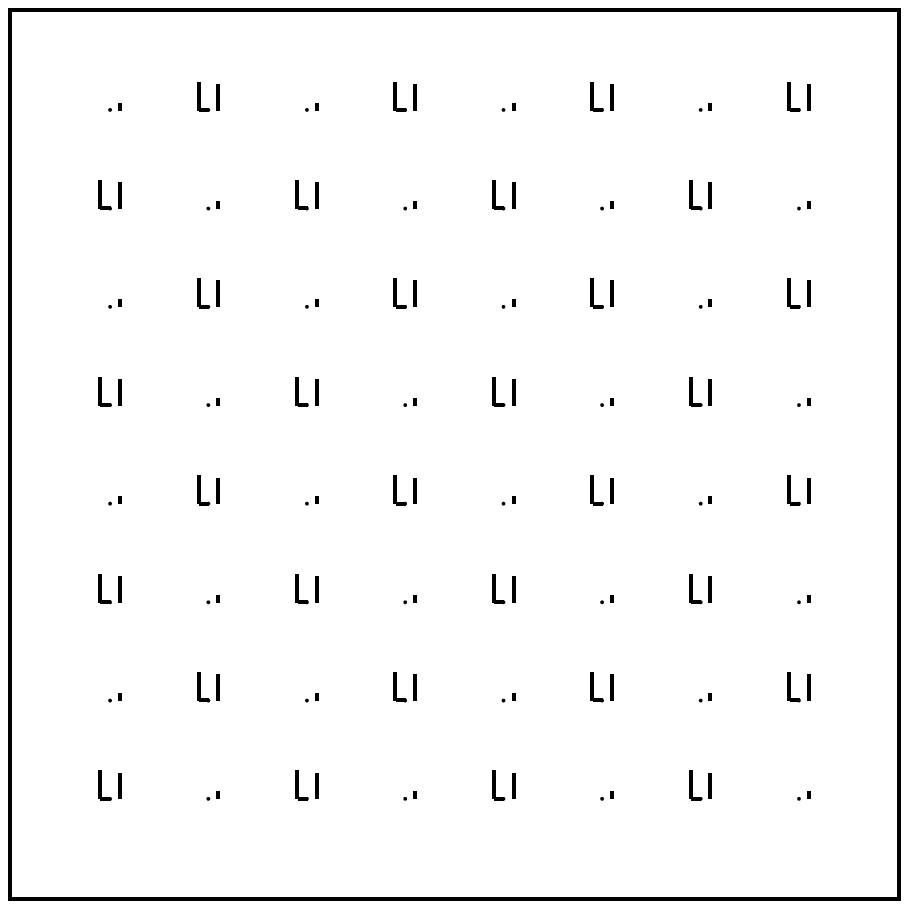}
}
\caption{%
Perfect checkerboard crystal-like order with two sublattices, as
obtained for $x=1/2$ doping in $8\times 8 $ cluster. The $C$-AF
spin order with alternating spin values between $S\simeq 1$ and
$S\simeq 1/2$ spins for charge majority/minority sites, and holes
doped into $x$ orbitals at charge minority sites. Uniform electron
distribution between $x$ and $z$ orbitals results in no JT
distortions at charge majority sites, while considerable JT
distortions at charge minority sites stabilize the orbital order
with (almost) empty $x$ orbitals. Meaning of symbols and the
values of parameters as in figure 1.
}
\end{figure}

\subsection{Maximal doping $x=1$ for LaSrNiO$_4$}

Experimental studies have established that for the examined class
of La$_{2-x}$Sr$_x$NiO$_4$ compounds with large doping $x\simeq
1$, the ground state is paramagnetic and conducting \cite{cava},
with a semiconductor-like conductivity (at low temperature close
to zero Kelvin) \cite{ivanova}. We investigated this case using
the present model by numerical computations, either using
parameter set of table 1 or somewhat modified parameters.
Extensive calculations have shown that reasonable values of
parameters do not give results in agreement with experimental
observations. In all considered cases one obtains perfect AF
order, with homogeneous charge distribution (one electron per
site) and with purely $z$-orbital occupation. In addition, the
HOMO-LUMO gap is quite large $\Delta=4.5$ eV (for the parameters
from table 1).

While this theoretical result is very reasonable and may be
expected taking the broken symmetry between two $e_g$ orbitals in
a monolayer LaSrNiO$_4$ system, it stimulates a question whether
this result could invalidate all the previous results reported
above. In our opinion the found disagreement of theory with
experiment for such large doping level is only an indication that
the effective model presented in this paper (without extra
modifications) is too limited to perform satisfactorily. Our
reasoning, based entirely upon experimental data, is twofold.

{\it First of all}, the conductivity at $x\simeq 1$ could be
likely due to a completely other mechanism \cite{ivanova}, namely
to sample quality and near impossibility to grow perfect
LaSrNiO$_4$ crystals. In reality what one grows are samples with
oxygen vacancies (at vertices of NiO$_6$ octahedra). The level of
oxygen vacancies is estimated to be at best $0.01-0.04$ which
seems enough to form a narrow donor band and therefore $p$-type
conduction could set in. How this $p$-band destroys the AF order
is another and difficult question. Still it is clear that the
two-band model is not appropriate for the description of
LaSrNiO$_4$.

{\it Second}, the model computations are based on a formal scheme:
the doping level is equal to fraction $x$ in the actual chemical
formula La$_{2-x}$Sr$_x$NiO$_4$. This is a common knowledge that
possibly this is not true, similar as in YBa$_2$Cu$_3$O$_{6+x}$
superconductors \cite{Zaa88}, and most probably this simple-minded
assumption works here reasonably well only for small values of
$x$. (Note that the same fault can be attributed to modelling done
in doped cuprates and manganites). Clearly, for larger doping one
can expect sizable deviations between electron doping level and
$x$ fraction in chemical formula. These deviations depend on the
type of the crystal lattice, the atom-atom distances, on the
constituent atoms and no simple formula exists to compute them.

We suggest that the question whether the first or second
explanation is more appropriate cannot be answered at present. It
could be that both are partly valid at the same time.

There are however some indications which can signal when and where
the above proportionality between doping level and the chemical
concentration $x$ could not be obeyed. In our particular case such
indications can be read out from ref. \cite{crystal06} where the
true quantum-chemistry {\it ab-initio} computations (HF, BLYP and
PBE) for LaNiO$_3$ were performed (using CRYSTAL06 computer code)
and where Mullikan population analysis was done. This is a
different (three-dimensional) conducting substance, not the one
studied in this paper, but still the problem is more or less the
same. The Mullikan charges found on Ni atoms are $\simeq +1.7$
instead of formal +3 and on oxygens the Mullikan charges are
$\simeq -1.2$ instead of formal -2. It is true that Mullikan
charges are only a rather crude estimate but still from the above
{\it ab-initio} data it follows that the effective model with
formal occupied with one $d$-electron-type Wannier function (per
site) is not correct for LaNiO$_3$.

One can envisage the modification of our model which possibly
could work properly for the description of LaSrNiO$_4$. It seems
necessary to supplement the model with at least one extra Wannier
orbital per site. For the first trial, in the spirit of the
Anderson lattice model, this could be a simple $s$-type Wannier
orbital of the oxygen type (assuming spherically symmetric
distribution of the oxygen $p$ electrons around the central
nickel). The new extra Hamiltonian terms would be simple kinetic
$s$-type nearest-neighbour hopping supplemented by $s-d$
hybridization. According to the Koster-Slater rules, the $s-d$
hybridization on single lattice site vanishes and one should
consider it between nearest-neighbour sites only. This goes
however beyond the scope of the present paper and is planned as a
separate future investigation.

\section{Crucial role played by Jahn-Teller distortions }

\subsection{Finite Jahn-Teller coupling $g_{\rm JT}$  }

We already remarked several times on the importance of JT coupling.
Here we try to draw some more general conclusions and provide extra
information.

Test computations for the undoped La$_2$NiO$_4$ substance ($x=0$)
invariably (independently of the initial conditions) give
insulating $G$-AF ground state (for $S=1$ spins) with uniform
charge distribution and equal occupation of $x$ and $z$ orbitals,
in perfect agreement with experiment \cite{tran08,buttrey86}. In
this state no JT distortions can arise. This result was obtained
for almost any set of the Hamiltonian parameters with one notable
exception. For too large JT coupling, in particular for $g_{\rm
JT}>3.5$ eV \AA${}^{-1}$ (while the other Hamiltonian parameters
are those from  the table 1) nonmagnetic phase starts to develop.
The emerging nonmagnetic sites feature equal occupation of 0.5
electron per up and down spin both for $x$ and $z$ orbitals. This
is accompanied by huge and clearly unphysical JT distortions of
the $Q_2$ type. For $g_{\rm JT}=3.8$ eV \AA${}^{-1}$ the common
magnetic sites with spin $S=1$ (and zero JT distortions) are
already quite scarce and the nonmagnetic sites do abound. We
suggest that for so large values of $g_{\rm JT}$ the 
simplified form of JT Hamiltonian which  we use is     
not proper for correct description of JT effects       
i.e., 
%
eq. (4), and in particular
the simple form of the coupling to the lattice  in eq. (4) 
requires substantial corrections.                     

\begin{figure}[t!]
\centerline{
\includegraphics[width=7.7cm]{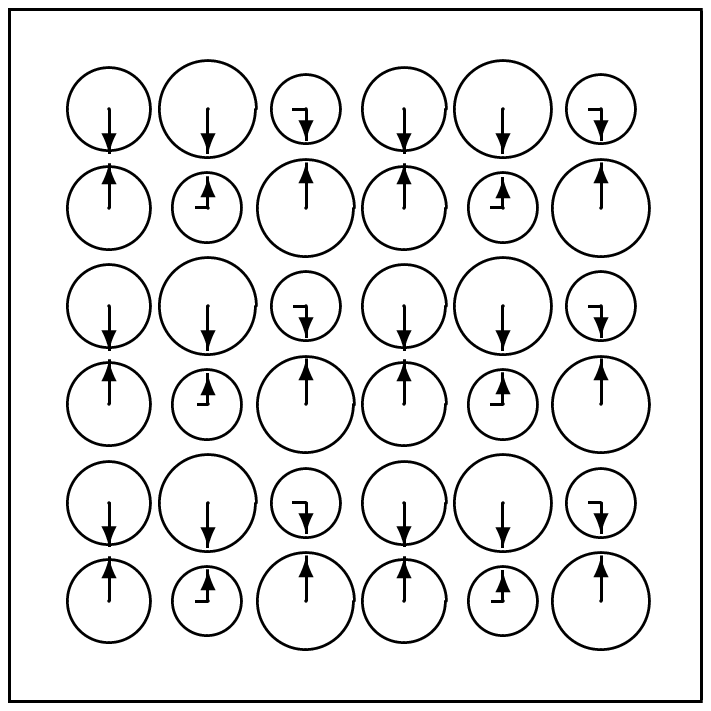}
}
\centerline{
\includegraphics[width=7.7cm]{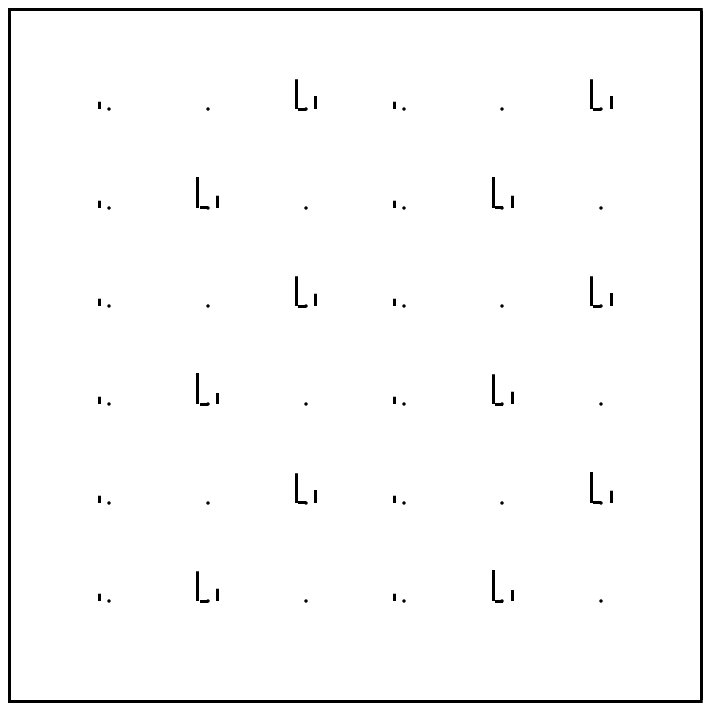}
}
\caption{%
Insulating ground state obtained for the $6\times 6$ cluster with
$x=1/3$ doping and zero crystal field ($E_z=0$). The $C$-AF order
develops, with spins $S=1$ and uniform charge distribution (the
same electron density in $x$ to $z$ orbitals) on undoped sites and
lower spins on charge minority sites.  The charge minority sites
do not form here diagonal stripe boundaries (as in figure 3) but
the superlattice with long-range order develops what is better
visible on the lower panel. Meaning of symbols and the values of
other parameters as in figure 1.
}
\end{figure}

Further, extensive tests for $x=1/3$ doped substance revealed that
only in a narrow region of $2.8<g_{\rm JT}<3.2$ (in eV\AA${}^{-1}$
units) diagonal stripes do form. Hence, the role of the JT
distortions is crucial for stripe formation. Furthermore, it was
also interesting to establish that for zero crystal field,
$E_z=0$, the diagonal stripes disappear, though the charge,
orbital and magnetic order remains almost the same like the one
for the standard parameters listed in table 1 (see figure 6,
compare with figure 3). The resulting phase is insulating, with
the HOMO-LUMO gap $\Delta=0.58$ eV.
(This is a somewhat similar scenario like the
one reported in Nd$_{1.67}$Sr$_{0.33}$NiO$_4$ \cite{hucker}).

However, when one switches to
smaller values of the on-site Coulomb repulsion, for example
taking $U=5$ eV and $E_z=0$, diagonal stripes reappear again
(i,e., when all the other parameters $\{t,J_H,g_{\rm JT},K\}$ have
the values given in table 1). The value $U=5$ eV corresponds to
suggestion that screening of
$U$ by hybridization with oxygen orbitals is larger than usually
assumed. It follows that the diagonal stripes do not appear for
any set (or even in a broad range) of the Hamiltonian parameters,
but result from a subtle balance between several competing
mechanisms. Furthermore, it is difficult to conclude what is more
proper for our nickelate model, either $U/t\approx 8$ or rather a
higher value $U/t\approx 12$ ($U/t\approx 12$ was our preference
for the bulk of the computations).

\subsection{Setting Jahn-Teller coupling to zero $g_{\rm JT}=$0.}

For zero JT coupling $g_{\rm JT}=0$ and small doping $x=1/8$ we detected
no change in the ground state --- it looks just like the one shown in
figure 1 (resulting from the computations performed for the standard
parameter set of table 1) and is insulating.

The situation  changes somewhat for higher doping. While for two
doping levels, $x=1/4$ and $x=3/8$, the ground states remain very
similar to the ones obtained before, i.e., the $C$-AF order is
accompanied by small charge modulation, when the JT distortions
are absent the $x=1/4$ state is metallic, opposite for the state
for doping $x=3/8$, where a relatively large gap $\Delta\simeq
0.7$ eV separates the HOMO and LUMO states in the $8\times 8$
cluster. For other larger dopings  (up to $x=1/2$) the ground
states are conducting.

This result, i.e., the exception obtained for $x=3/8$, is puzzling
and could be either a finite size effect and/or an indication of
quite different electron correlation energies in systems with
different electron number. Here it seems that a full explanation
and a few digressions are proper. To be precise we performed
control computations in $8\times 8$ cluster doped with 26 holes
(quite close to 24 holes which correspond to $x=3/8$).
The true ground state was conducting.
Then we performed another set of control computations
in $6\times 6$ cluster doped with 14 holes.
(This is as close as possible to  $3/8$ doping in $8 \times 8$
cluster). This time the situation was a little bit more complicated.
The best HF ground state (charge homogeneous with
well developed $C$-AF order) was found to be insulating. The
corresponding correlation energy correction was rather
small. However, for a group of higher HF energy levels (with
a similar but non-perfect ordering and with HOMO-LUMO gap being
about 0.1 eV, i.e., possibly conducting) much stronger
correlations develop and when included the total energy turned out
to be lower than that obtained for the first HF ground state
plus its correlation energy. In other words for $6\times 6$
and for doping $x \approx \frac{3}{8}$ the true
ground state is conducting and strong correlations are
responsible for that. Why the same situation was not found for
$8\times 8$ cluster can, possibly, be due to poor selection of the
local operators (within local ansatz), see eq.~\ref{om} which
could be thus responsible for not catching some sizable part of
total correlation energy. In other words mean-field type treatment of
correlations (which we employ)
could be not enough, and the true physical picture is
that various contributions  to correlation energy are highly
non-homogeneous and strong in the doped, charge and spin
non-homogeneous, nickelates.

As the closing remarks to this lengthy digression let us mention
that similar control computations were performed in $6\times6$
clusters for different dopings --- the results corroborated the
results obtained in $8\times 8$ clusters (the $x =\frac{3}{8}$
doping was the only exception).

Now, let us return back to the main subject and other doping levels.
For $x=1/2$ the ground state is again an experimental
checkerboard: one finds a two-sublattice arrangement, exactly like
the one displayed in the top panel of figure 5. The long range
charge and orbital order are both close to being perfect here.
This demonstrates that the checkerboard phase can form by a purely
electronic mechanism, and the JT distortions only make this state
more robust. (For zero JT coupling this state is also conducting).

As mentioned before, for doping $x=1/3$, one obtains a conducting
ground state with $g_{\rm JT}=0$, i.e., the HOMO-LUMO gap is
$\Delta=0.03$ eV. The obtained phase shows a slightly
non-homogeneous $C$-AF order which looks very similar to the upper
panel on figure 2. The true long-range order is however not
present and diagonal stripes do not appear.

\section{Summary and conclusions}

Let us address the correlated wave functions first.
As expected the correlation effects turned out to be much stronger here
than in the manganites \cite{rosc}. This could be recognized by
investigating nonmagnetic states, where the energy is dramatically
reduced by the correlation effects (not shown). In the states with
either $G$-AF (for $x=0$) or $C$-AF order (for $0<x\le 0.5$) the
correlation energy is rather small as the magnetic states are well
developed. This demonstrates that the Hartree-Fock approximation
provides a realistic energy estimate for the states with broken
symmetry when the Coulomb interactions are strong. A similar situation
was also found in the two-band model for superconducting cuprates
\cite{Ole89}.

Coming to detailed analysis, the obtained energies of the ground
state found in the Hartree-Fock approximation and using the local
ansatz are shown in table 2. In each case the correlation energy,
$E_{\rm corr}=E_{\rm HF}-E_{\rm tot}$ is rather small. The
insulating gap exhibits a non-monotonous behavior: it is large for
the undoped La$_2$NiO$_4$ compound ($x=0$), with a HOMO-LUMO gap
of 7.42 eV (see table 2). It drops to 1.86 eV for $x=0.125$, and
becomes rather small for $x=0.25$, where we found a local minimum
of $0.28$ eV. In case of the stripe phase at $x=1/3$ it is again
larger (note that the size of the cluster $6\times 6$ is different
here, but this should not influence significantly the estimated
gap $\Delta$), and next increases steadily up to $\Delta=0.91$ eV
at $x=0.5$. This confirms the experimental observation that
La$_{2-x}$Sr$_x$NiO$_4$ nickelates are insulators \cite{tran08} in
the entire doping regime up to half-doping $x=1/2$.

\begin{table}[t!]
\caption{\label{tab:corr} Energies obtained for the insulating ground
states in the Hartree-Fock approximation, $E_{\rm HF}$, and for
correlated wave functions, $E_{\rm tot}$, as well as the HOMO-LUMO gap
$\Delta$ obtained for representative doping levels (all in eV).
Last row shows the prediction of the model for LaSrNiO$_4$ ($x=1$) ---
this result does not correlate well with experiment, see section 3.4.
}
\begin{center}
\begin{tabular}{ccrrcc}
\hline
  $x$   & cluster & $E_{\rm HF}$ & $E_{\rm tot}$ & $\Delta$  \\
\hline
  0.000 & $8\times 8$ & 263.64 & 263.54  &  7.42&  \\
  0.125 & $8\times 8$ & 221.87 & 221.63  &  1.86&  \\
  0.250 & $8\times 8$ & 179.95 & 178.90  &  0.28&  \\
  0.333 & $6\times 6$ &  85.14 &  84.31  &  0.57&  \\
  0.375 & $8\times 8$ & 136.08 & 134.96  &  0.72&  \\
  0.500 & $8\times 8$ &  95.14 &  93.29  &  0.91&  \\ \hline
  1.000 & $8\times 8$ & -46.17 & -46.18  &  4.52&  \\ \hline
\end{tabular}
\end{center}
\end{table}

The role of JT distortions for the phase situation in
La$_{2-x}$Sr$_x$NiO$_4$ was discussed in the literature
\cite{theo1,theo2}, but the results are controversial to some extent.
In ref. \cite{theo1} the stripes were attributed to Jahn-Teller
distortions but
in ref. \cite{theo2} it was argued (for the model explicitly featuring
oxygens in-between nickel atoms) that Jahn-Teller distortions are not
consistent with the local symmetry and with the experimental results.
Our simple effective model is not able to resolve this controversy and
to answer the question concerning the origin of the stripe phase found
at $x=1/3$ doping as it considers distortions as being independent from
one another (i.e., not cooperative). Still within the framework of the
present model we support the claim by Hotta and Dagotto \cite{theo1}
that Jahn-Teller distortions are essential for the development of the
stripe order for the doping $x=1/3$. On the other hand, other results
of our computations show that the short range order (on average) does
not change due to the lattice distortions, and it is only the global
population of $3z^2-r^2$ orbitals at defect sites and long range order
which are stabilized by the Jahn-Teller effect.

The results of the present study uncover the fundamental
difference between the stripes in cuprates and in nickelates.
While a competition between the magnetic energy (superexchange)
and the kinetic energy of doped holes drives the stripe structures
in doped cuprates \cite{strip} with site-centered structures
\cite{Tra08,Gre10}, or could also be responsible for stripes in
certain $t_{2g}$ systems \cite{Wro10}, this mechanism is absent in
layered nickelates for two reasons: ($i$) when hole doping occurs
in $d^8$ (instead of $d^9$) configuration, the ions occupied by
holes carry a spin $1/2$ of the remaining $e_g$ electron and
participate in the magnetic order, and ($ii$) the doped ions are
active for the Jahn-Teller effect and local distortions form
around them. Both these effects suppress the kinetic energy of the
doped holes and the competition between the magnetic and kinetic
energy is absent. Therefore, the stripes cannot form at low doping
(up to $x=0.25$ considered here) but this regime is dominated by
isolated charge defects stabilized by local lattice distortions.
The doping $x=1/3$ is the first one where the doped sites have to
appear as next nearest neighbours, and they order in form of a
stripe phase. We suggest that the electronic structure has a local
minimum for this phase rather than for a random distribution of
doped sites. This pattern does not hold at higher doping
$x=0.375$, but decides about the stability of the checkerboard
phase at the doping $x=0.5$, where again the configurations with
doped ions being next nearest neighbours to one another occur along
both diagonal directions. Altogether, the diagonal stripe pattern
for $x=1/3$ is thus favoured by the characteristic properties of
doped holes in layered nickelates which avoid each other.

Summarizing, the main virtue (in our opinion) of this paper is a
demonstration that the same model which worked very well for doped
perovskite manganites \cite{rosc} performs equally well for the
layered nickelates. It helped to identify the leading mechanism,
being the Jahn-Teller effect in presence of strong local
correlations, responsible for the stripe phase which occurs only
at the $x=1/3$ doping and for the checkerboard phase found at the
$x=1/2$ doping. At the same time some open questions still remain
which could be resolved by future studies. In particular, it would
be interesting to perform a theoretical study with cooperative
Jahn-Teller distortions including explicitly oxygen ions.
Furthermore, it would be very helpful in future theoretical
modeling of doped nickelates if a more precise relation between
the electron doping level (used in computations) and the chemical
doping (such as follows from the chemical formula) could be
established.

\ack

We acknowledge financial support by the Polish Ministry of Science
and Higher Education under Project No. N202~104138.
A M Ole\'s was also supported by the Foundation for Polish Science (FNP).

\end{document}